\documentclass[aps,pra,twocolumn,superscriptaddress]{revtex4-1}
\usepackage[utf8]{inputenc}
\usepackage{graphicx}
\usepackage{amsmath}
\usepackage[amssymb,Gray]{SIunits}
\usepackage[caption=false]{subfig}

\newcommand{\bra}[1]{\langle #1 \hspace{-2pt} \mid}
\newcommand{\ket}[1]{\mid \hspace{-1pt} #1 \rangle}
\newcommand{\erw}[1]{\ensuremath{\langle #1 \rangle}}
\newcommand{\abs}[1]{\ensuremath{\left\vert #1 \right\vert}}
\newcommand{\D}{\mathrm{d}}

\renewcommand{\vec}[1]{\boldsymbol{\mathrm{#1}}}

\begin{document}


\title{On approximations for the free evolution of self-gravitating quantum particles}


%
\author{Andr\'e Gro{\ss}ardt}
\email[]{andre.grossardt@ts.infn.it}
\affiliation{Department of Physics, University of Trieste, 34151 Miramare-Trieste, Italy}
\affiliation{Istituto Nazionale di Fisica Nucleare, Sezione di Trieste, Via Valerio 2, 34127 Trieste, Italy}


\date{\today}

\begin{abstract}
The evolution of the centre-of-mass wave-function for a mesoscopic particle
according to the Schr{\"o}\-din\-ger-New\-ton equation can be approximated by a harmonic potential, if the wave-function is narrow compared to
the size of the mesoscopic particle. It was noticed by Colin et~al. [Phys. Rev. A, \textbf{93}, 062102 (2016)] that, in the regime where self-gravitational effects are weak,
intermediate and wider wave-functions may be approximated by a harmonic potential as well, but with a width dependent coupling, leading to a time evolution that is determined only by a differential equation for the width of
a Gaussian wave-function as a single parameter. Such an approximation results in considerably less computational effort in order to predict the self-gravitational effects on the
wave-function dynamics.
Here, we provide an alternative approach to this kind of approximation, including a rigorous derivation of the equations of
motion for an initially Gaussian wave packet, under the assumption that its shape is conserved. Our result deviates to
some degree from the result by Colin et~al., specifically in the limit of wide wave-functions.
\end{abstract}


\maketitle



\section{Introduction}

As long as there is no agreement on what the correct theory is that unifies quantum physics
and gravitation, there is no unambiguous prediction for the gravitational interaction of
quantum matter. While most physicists prefer the idea that the gravitational field itself
must exhibit quantum properties at the microscopic regime, a fundamentally semi-classical coupling of
quantum matter to classical space-time has been proposed as an
alternative~\cite{Moller:1962,Rosenfeld:1963,Carlip:2008}. A natural way to source the curvature of
classical space-time by quantum matter is provided by the semi-classical Einstein equations~\cite{Moller:1962,Rosenfeld:1963},
\begin{equation}
\label{eqn:sce}
R_{\mu \nu} - \frac{1}{2} g_{\mu \nu} R = \frac{8 \pi G}{c^4} \,
\bra{\Psi} \hat{T}_{\mu \nu} \ket{\Psi} \,,
\end{equation}
where the stress-energy-tensor of the classical Einstein equations is replaced by the expectation value of
the corresponding quantum operator. While some controversies exist about the relevance of this
approach~\cite{Eppley:1977,Page:1981}, the arguments against it are
inconclusive~\cite{Mattingly:2005,Mattingly:2006,Albers:2008}. The ultimate decision whether or not the gravitational field
must be quantised, therefore, seems to be up
to experiments~\cite{Rosenfeld:1963,Mattingly:2005,Carlip:2008}.

Interestingly, the mass regime where such experiments become feasible is not that far from what is currently possible in the laboratory.
This is because eq.~\eqref{eqn:sce} yields a nonlinear Newtonian gravitational self-interaction in the nonrelativistic Schr{\"o}\-din\-ger
equation. Such a nonlinearity has observable consequences, despite the weakness of the gravitational interaction.

These self-gravitational effects were first modelled by the \emph{one-particle} Schr{\"o}\-din\-ger-New\-ton equation~\cite{Carlip:2008,Giulini:2011},
\begin{multline}
\label{eqn:sn}
\mathrm{i} \hbar \frac{\partial}{\partial t} \psi(t,\vec r) = \Bigg( -\frac{\hbar^2}{2 m} \nabla^2 \\
- G m^2 \int \D^3 \vec r' \, \frac{\abs{\psi(t,\vec r')}^2}{\abs{\vec r - \vec r'}}
\Bigg) \psi(t,\vec r) \,.
\end{multline}
The self-gravitational interaction represented by the nonlinear term predicts inhibitions of the free spreading of a
wave-function for masses of about $\unit{10^{10}}{\atomicmass}$ and beyond~\cite{Giulini:2011}. Experiments aiming at
testing this behaviour could be realistic, at least in space, where a free spreading over longer time scales is feasible.

Evidently, the massive systems needed for such an experiment cannot be elementary particles. Rather will they consist
of a large number of constituents. One may, therefore, ask in what sense eq.~\eqref{eqn:sn} describes such a many-body system.
The many-body dynamics following from the hypothesis of a fundamentally semi-classical gravity~\eqref{eqn:sce}
will be discussed in some detail in section~\ref{sec:manypart}.
One finds that---quite contrary to the situation where it is considered as a Hartree
approximation~\cite{Adler:2007}---eq.~\eqref{eqn:sn} approximately describes the centre of mass of a complex system,
provided that its centre of mass is sufficiently de-localised in order to be allowed to disregard its internal structure.

On the other hand, if the centre-of-mass wave-function is narrow compared to the size of the system, the resulting self-gravitational
potential is a function of the first and second moments, $\erw{\vec r}$ and $\erw{r^2}$, and quadratic in $\vec r$.
A potential of this form will leave the shape of a Gaussian wave-function intact during its evolution.
Contrary to eq.~\eqref{eqn:sn}, where for a numerical analysis the full wave-function must be simulated,
the centre-of-mass dynamics can then be described by the evolution of the wave-function width as a single parameter.

Colin et.~al~\cite{Colin:2016} make use of this, proposing to approximate the Schr{\"o}\-din\-ger-New\-ton equation
by a quadratic potential also in the case where the full dynamics are given by eq.~\eqref{eqn:sn}.
Their approximation is valid for weak gravitational potentials and substantially simplifies numerical calculations.
It has been employed in order to
provide a crucial contribution towards an experimental test by comparing the self-gravitational inhibitions
of dispersion to other decoherence sources.

The wave-function dependent quadratic potential they use was, however, inferred from a comparison
of energies, rather than derived directly from the Schr{\"o}\-din\-ger-New\-ton equation. Here, after reviewing the
many-body and centre of mass equations in section~\ref{sec:manypart}, we will provide such a direct derivation in
section~\ref{sec:evolution}. The result will be compared with the approximation used by
Colin et~al.~\cite{Colin:2016,Colin:2014a} in section~\ref{sec:comparison},
where we find some deviations. We will conclude with a brief discussion.

\section{The many-body Schr{\"o}\-din\-ger-New\-ton equation}\label{sec:manypart}

In a condensate of many constituents which collectively occupy the same state, eq.~\eqref{eqn:sn} appears as
a Hartree approximation and can be derived in a similar fashion as the Gross-Pitaevskii
equation~\cite{Gross:1961,Pitaevskii:1961,Bardos:2002}; it is then, for example,
a model for the self-gravitation of Bosonic stars~\cite{Ruffini:1969}.
In this context, its origin is the \emph{mutual} Newtonian interaction of the constituents and it provides an effective
description of the factors of a tensor product state of the condensate, regardless whether eq.~\eqref{eqn:sce} is fundamentally
correct or not.
This is \emph{not} the point of view taken here.

On the other hand, if one starts with eq.~\eqref{eqn:sce} as a hypothesis about the gravitational interaction of quantum matter,
applies it to a Klein-Gordon or Dirac field, takes the nonrelativistic limit, and restricts oneself to the one-particle
sector then eq.~\eqref{eqn:sn} follows immediately as the \emph{fundamental} dynamical equation for a \emph{single}, nonrelativistic
quantum particle~\cite{Giulini:2012,Bahrami:2014}. Since in the nonrelativistic limit there is no interaction between the sectors of
different particle number, one can choose a restriction to the $N$-particle sector instead, for arbitrary $N$. In contrast to the
Hartree approximation which is already an effective description for $N$ constituents, one then obtains a \emph{different} equation for
the dynamics of the full $N$-particle wave-function~\cite{Diosi:1984,Bahrami:2014}:\vspace{-5pt}
\begin{subequations}\begin{multline}\label{eqn:n-particle-sn}
\mathrm{i} \hbar \frac{\partial}{\partial t} \Psi(t,\vec r_1,\cdots,\vec r_N)
= \biggl(-\sum_{i=1}^N\frac{\hbar^2}{2m_i}\nabla^2_i \\
+ V_\text{nongrav.} +  V_g[\Psi] \biggr)\Psi(t;\vec r_1,\cdots \vec r_N) \,
\end{multline}
with the gravitational potential\vspace{-5pt}
\begin{multline}
V_g[\Psi](t,\vec r_1,\cdots,\vec r_N) = -G\sum_{i=1}^N\sum_{j=1}^N m_i m_j \\
\times \int \D^3 r_1' \cdots \D^3 r_N'
\frac{\abs{\Psi(t;\vec r'_1,\cdots,\vec r'_N)}^2}{\abs{\vec r_i - \vec r'_j}} \,.
\end{multline}\end{subequations}
$\vec r_i$ and $m_i$ are the coordinates and masses of the $N$ constituents.
The gravitational potential $V_g$ contains both the self-gravitational interaction of each constituent and the mutual
gravitational interactions between different constituents.

Due to the nonlinear gravitational term, eq.~\eqref{eqn:n-particle-sn} does not strictly separate into centre of mass
and relative coordinates. Nonetheless, one can apply an appropriate, Born-Oppenheimer-type, approximation scheme,
making use of the fact that the gravitational interaction is much weaker than intramolecular forces, which are making
up the internal structure of the system modelled by a mass density distribution $\rho_c$. Precisely speaking, if the
wave-function separates into centre of mass and relative coordinates,
$\Psi(t;\vec r_1,\cdots,\vec r_N) = \psi(t,\vec r) \chi(\vec r_1,\dots,\vec r_{N-1})$,
then $\rho_c$ is given by the relative wave-function~\cite{Giulini:2014}:
\begin{multline}
\rho_c(\vec r)
=\sum_{i=1}^{N-1} m_i \int \D^3 r_1 \cdots \D^3 r_{i-1} \, \D^3 r_{i+1} \cdots \D^3 r_{N-1} \\
\times \abs{\chi(\vec r_1,\dots,\vec r_{i-1},\vec r,\vec r_{i+1},\dots,\vec r_{N-1})}^2\,.
\end{multline}
The centre-of-mass wave-function then satisfies~\cite{Giulini:2014}
\begin{subequations}\label{eqn:sn-born-oppen}\begin{align}
\mathrm{i} \hbar \frac{\partial}{\partial t} \psi(t,\vec r) &= \left( -\frac{\hbar^2}{2 m} \nabla^2
+ V_g[\psi](t,\vec r) \right) \psi(t,\vec r) \\
V_g[\psi](t,\vec r) &= -G \int \D^3 \vec r' \;\abs{\psi(t,\vec r')}^2 \, I_{\rho_c}(\abs{\vec r - \vec r'}) \\
I_{\rho_c}(\vec d) &= \int \D^3\vec u \int \D^3\vec v\;
\frac{\rho_c(\vec u)\rho_c(\vec v-\vec d)}{\abs{\vec u - \vec v}}  \,.
\end{align}\end{subequations}
Here $\vec r$ is the centre-of-mass coordinate, $\psi$ the centre-of-mass wave-function, and
$\rho_c$ the mass density of the many-body system with respect to its centre of mass.
The function $I_{\rho_c}$ is proportional to the gravitational energy between a mass distribution $\rho_c$ and
a copy of the same mass distribution shifted by a distance $\vec d$.
Equation~\eqref{eqn:sn-born-oppen} is the Schr{\"o}\-din\-ger-New\-ton equation for the centre of mass
of a complex system of many constituents. This is the equation we are interested in.

If the system is modelled by a homogeneous, spherical mass distribution, the potential $I_{\rho_c}$ takes the form~\cite{Iwe:1982}
\begin{multline}\label{eqn:self-energy-homogeneous-sphere}
I_{\rho_c}^\text{sphere}(d)=\frac{m^2}{R} \\ \times
\begin{cases}
\frac{6}{5}-2\left(\frac{d}{2R}\right)^2+\frac{3}{2}\left(\frac{d}{2R}\right)^3-\frac{1}{5}\left(\frac{d}{2R}\right)^5
& (d\leq 2R)\,,\\
\frac{R}{d}
& (d > 2R)\,.
\end{cases}
\end{multline}
One can see immediately that in situations where the mass distribution is point-like compared to the
wave-function, eq.~\eqref{eqn:sn-born-oppen} is well-approximated by eq.~\eqref{eqn:sn}.

In the opposite case, where the spatial wave-function can be considered narrow, and the mass is distributed
homogeneously over a sphere of radius $R$, a good approximation to $V_g$ is given by~\cite{Yang:2013,Giulini:2014}
\begin{multline}\label{eqn:pot-narrow-sphere}
V^\text{sphere}_\text{narrow}[\psi](t,\vec r) = \mathrm{const.} \\
+ \frac{m \omega_\text{SN}^2}{2} \Big( \vec r^2
- 2 \vec r \cdot \bra{\psi} \vec r \ket{\psi}
+ \bra{\psi} \vec r^2 \ket{\psi} \Big) \,.
\end{multline}
For a mesoscopic, homogeneous sphere one finds $\omega_\text{SN}^2 = G m / R^3$.
The nonlinearity is still present, encoded in the expectation values.
Based on this approximation, tests of the Schr{\"o}\-din\-ger-New\-ton equation
have been proposed also with optomechanical experiments~\cite{Yang:2013,Grossardt:2015b,Grossardt:2016}.

Since the potential~\eqref{eqn:pot-narrow-sphere} depends
only on the first and second moments and is quadratic in $\vec r$, it can be shown to leave the shape of an
initially Gaussian wave packet unchanged~\cite{Yang:2013,Colin:2014a,Colin:2016}.
Therefore, the full (as far as the narrow wave-function approximation is valid)
dynamics for a Gaussian wave packet are given by the time evolution of the width $\sqrt{\erw{r^2}}$ of the wave packet.

\section{Evolution of a self-gravitating Gaussian wave-function}\label{sec:evolution}
For wider wave-functions, the gravitational potential deviates from the quadratic form~\eqref{eqn:pot-narrow-sphere},
and the wave-function shape will change over time.
If the self-gravitational interaction is weak (i.\,e. the kinetic energy of the mesoscopic particle is much larger than the
gravitational energy), one may, however, assume as a first order approximation that the shape of the wave-function
stays a Gaussian, and self-gravitation only affects its width~\cite{Colin:2016,Colin:2014a}.

In the case of a spherically symmetric wave-function, the expectation values of radial momentum and distance vanish exactly,
$\erw{r} = \erw{p} = 0$. The equations of motion for the second moment $u(t) = \erw{r^2}(t)$ follow straightforwardly from the
Schr{\"o}\-din\-ger-New\-ton equation~\eqref{eqn:sn-born-oppen}. One finds
\begin{subequations}\begin{align}
\frac{\partial}{\partial t} \erw{r^2}(t) &= \frac{1}{m} \erw{\vec r \cdot \vec p + \vec p \cdot \vec r}(t) \\
\frac{\partial}{\partial t} \erw{p^2}(t) &= - \erw{\vec p \cdot (\nabla V_g) + (\nabla V_g) \cdot \vec p} \nonumber\\
&= -m \frac{\partial}{\partial t} \erw{V_g}(t) \label{eqn:psquared}\\
\frac{\partial}{\partial t} \erw{\vec r \cdot \vec p + \vec p \cdot \vec r}(t) &=
\frac{2}{m} \erw{p^2}(t) - \erw{\vec r \cdot \nabla V_g}(t) \,,
\end{align}\end{subequations}
where the second equality in eq.~\eqref{eqn:psquared} is shown in ref.~\cite{Grossardt:2015b}.
If now one applies the aforementioned approximation, assuming that the gravitational potential $V_g$ is simply a function
of the width $u(t)$ of a Gaussian state, this yields
\vspace{-5pt}
\begin{subequations}\begin{equation}\label{eqn:diffeq}
\dddot{u}(t) = -3 \omega_\text{SN}^2 \, f(u(t)) \, \dot{u}(t) \vspace{5pt}
\end{equation}
with (see the Appendix for a more detailed derivation)
\begin{align}
f(u) &= \frac{\partial}{\partial u} F(u) \\
F(u) &= \frac{R^3}{3 G m^2} \erw{2 V_g + \vec r \cdot \nabla V_g} \nonumber\\
&= -\frac{144 \,R^8}{\pi \, u^3} \int_0^\infty \D p \, \int_0^p \D q \,
\mathrm{e}^{-\frac{3 R^2}{u} (p^2 + q^2)} \nonumber\\
&\phantom{=}\; \times (p^2 - q^2) \, \left(\Pi(p) - \Pi(q)\right)\,.
\end{align}
For a homogeneous sphere mass distribution, the function $\Pi$ takes the form
\begin{equation}
\Pi(x) = \begin{cases}
\frac{6}{5}x^2-\frac{3}{2}x^4+\frac{21}{20}x^5-\frac{9}{70}x^7
& (x\leq 1)\,,\\
\frac{3}{4}x
& (x > 1)\,.
\end{cases}
\end{equation}\end{subequations}
Note that the dependence on mass and the gravitational constant are completely encoded in $\omega_\text{SN}$,
and the function $f$ depends only on the wave-function width $u$, as well as the radius $R$ of the mesoscopic particle.
For the homogeneous sphere, the integral can be solved analytically, resulting in
\begin{multline}
f(u) = \mathrm{erf}\left(\sqrt{\frac{3}{u}}\right) +\sqrt{\frac{u}{3\pi}} \\
\times \left(u-\frac{7}{2}-\frac{324 - 162 u - 35 u^4+ 70 u^5}{70 \, u^4}\, \mathrm{e}^{-3/u}\right) \,,
\end{multline}
where $\mathrm{erf}$ is the Gauss error function and $u$ is expressed in units of $R^2$.
One can easily check that this function has the appropriate limits
\begin{equation}
\lim_{u \to 0} f(u) = 1 \, \quad \text{and} \quad \lim_{u \to \infty} f(u) = 0 \,.
\end{equation}

\begin{figure*}
\centering
\includegraphics{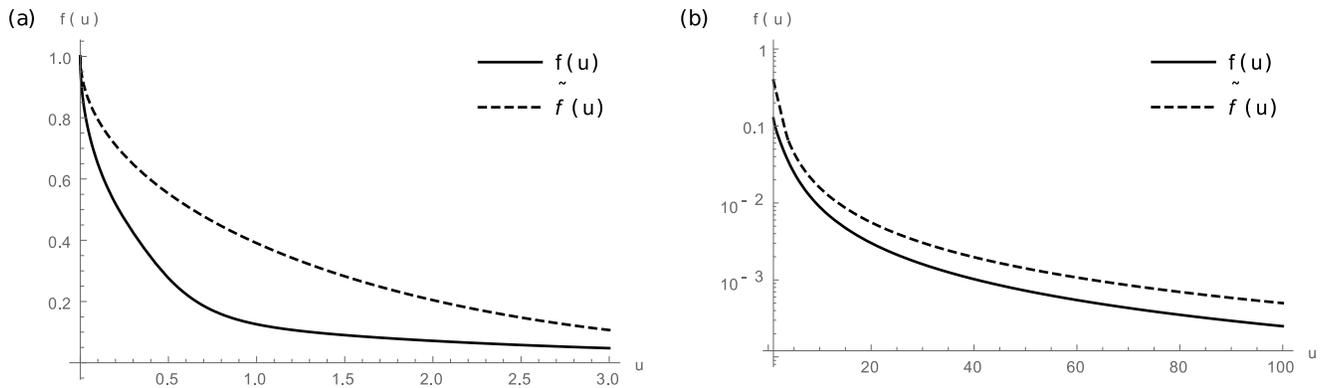}
\caption{Comparison of the function $f(u)$ defined in eq.~\eqref{eqn:diffeq} (solid curve) and the function $\widetilde{f}(u)$ 
defined in eq.~\eqref{eqn:f-g-colin} (dashed curve). $u$ is in units of $R^2$. (a) for small $u$. (b) for large $u$ (logarithmic scale).}
\label{fig:plot-functions}
\end{figure*}

\subsection{Initial conditions}
For an initial real valued Gaussian of width $\erw{r^2}_0 = u_0$,
\begin{equation}
\psi(t=0,r) = \left(\frac{3}{2\pi u_0}\right)^{3/4} \, \exp\left(-\frac{3 r^2}{4 u_0}\right) \,,
\end{equation}
the differential equation~\eqref{eqn:diffeq} must be solved with initial conditions
\begin{subequations}\begin{align}
u(0) &= u_0 \\
\dot{u}(0) &= 0 \\
\ddot{u}(0) &= \frac{9 \hbar^2}{2 m^2 \, u_0} - \omega_\text{SN}^2\,g(u_0) \,u_0 \label{eqn:ini-uss}
\end{align}\end{subequations}
with the function
\begin{align}
g(u) &= \frac{R^3}{G\,m^2\,u} \erw{\vec r \cdot \nabla V_g}_0 \nonumber\\
&= -\frac{432\,R^8}{\pi \, u^4} \,\int_0^\infty \D p \, \int_0^p \D q \,
\mathrm{e}^{-\frac{3 R^2}{u} (p^2 + q^2)} \nonumber\\
&\phantom{=}\; \times (p^2 - q^2) \, \left(K(p) - K(q)\right)\,.
\end{align}
For the homogeneous sphere, $K$ is
\begin{equation}
K(x) = \begin{cases}
-\frac{1}{2}x^4+\frac{9}{20}x^5-\frac{1}{14}x^7
& (x\leq 1)\,,\\
-\frac{1}{4}x
& (x > 1)\,,
\end{cases}
\end{equation}
which yields (again for units with $R=1$)
\begin{multline}
g(u) = \mathrm{erf}\left(\sqrt{\frac{3}{u}}\right) + \sqrt{\frac{u}{3 \pi }} \\
\times \left(\frac{2}{3} u - 3 +\frac{486+105 u^3-70 u^4}{105 u^3}\, \mathrm{e}^{-3/u}\right) \,.
\end{multline}
The dynamics depend, in general, on the radius $R$ of the mesoscopic sphere, its mass $m$, and the initial width $u_0$
of the centre-of-mass wave-function. However, the
dependence on the latter is only a dependence on the ratio $u_0/R$ for the functions $f$ and $g$,
while $\omega_\text{SN}$ is a function of mass density only. For a given mass density, if units are chosen such that $R=1$,
the only true dependence on mass comes through the initial condition~\eqref{eqn:ini-uss}.

While for a precise numerical analysis of the equations~\eqref{eqn:sn} and~\eqref{eqn:sn-born-oppen} the full wave-function must be
simulated on a grid of many points, for this approximation it is sufficient to determine the evolution
of $u(t) = \erw{r^2}$. This significantly simplifies numerical estimations of the magnitude of effects, and may even allow for
some analytical results which cannot be seen from the full equation.

\section{Comparison with previous results}\label{sec:comparison}

\begin{figure*}
\centering
\includegraphics{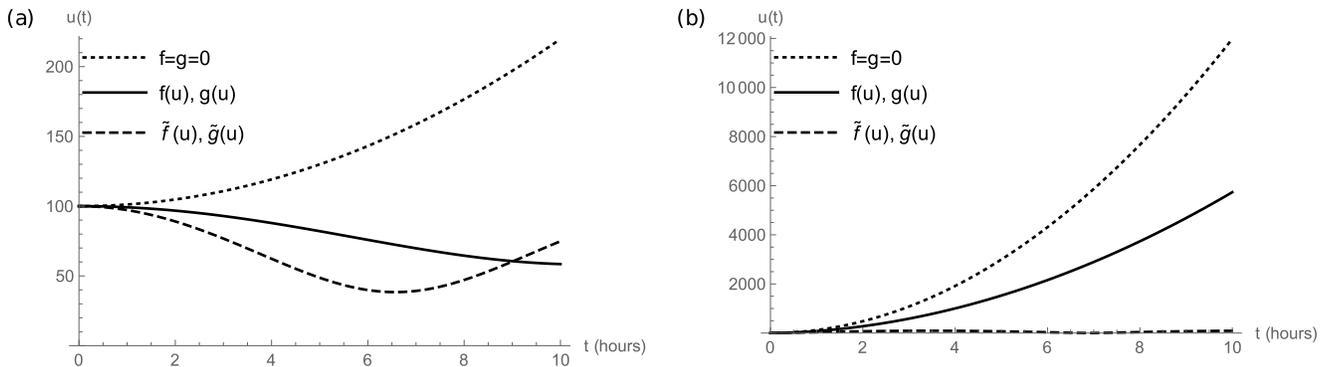}
\caption{Solution of the equations of motion for a wide and intermediate wave-function.
The dotted curve shows the free solution in absence of self-gravity, the solid curve is the solution using
$f(u)$ and $g(u)$ as defined in eqs.~\eqref{eqn:diffeq} and~\eqref{eqn:ini-uss}, and the dashed curve uses the approximation
according to eq.~\eqref{eqn:f-g-colin}.
The mass is chosen to be $m = \unit{10^{10}}{\atomicmass}$ for an Osmium sphere ($\rho = \unit{22.6}{\gram\per\centi\meter\cubed}$),
corresponding to a diameter of approx. \unit{100}{\nano\meter}.  $u$ is in units of $R^2$.
(a) wide wave-function, $u_0 = (10 R)^2$. (b) intermediate wave-function, $u_0 = R^2$.}
\label{fig:plot-solution}
\end{figure*}

Equation~\eqref{eqn:diffeq} also follows, if one starts with a harmonic oscillator ansatz with width dependent coupling,
\begin{equation}\label{eqn:sg-quad}
\mathrm{i} \hbar \frac{\partial}{\partial t} \psi(t,\vec r) = \left( -\frac{\hbar^2}{2 m} \nabla^2
+ \frac{k(u)}{2} r^2 \right) \psi(t,\vec r) \,.
\end{equation}
The authors of~\cite{Colin:2016,Colin:2014a} employ an energy argument in order to determine an approximation
for $k(u)$. To follow this argument, first note that equation~\eqref{eqn:sg-quad} possesses the conserved energy
\begin{equation}
E = \frac{\hbar^2}{2 \,m} \, \int \D^3 \vec{x} \, \abs{\nabla \psi(t,r)}^2 + U(u(t)) \,,
\end{equation}
if the potential $U$ satisfies
\begin{equation}
\frac{\D U}{\D u} = \frac{k(u)}{2} \,.
\end{equation}
For the standard harmonic oscillator potential, where $k$ is a constant, $U$ is simply the potential energy
of the oscillator.
The potential $U$ is then chosen such that $U(u) = -G I_{\rho_c}^\text{sphere}(\sqrt{u})$, with $I_{\rho_c}^\text{sphere}$ defined
in eq.~\eqref{eqn:self-energy-homogeneous-sphere}, which leads to the functions
\begin{subequations}\label{eqn:f-g-colin}\begin{align}
\widetilde{f}(u) &= \begin{cases}
1 - \frac{21}{32} \sqrt{u} + \frac{3}{64} \sqrt{u^3} & (u \leq 4) \\
\frac{1}{2\sqrt{u^3}} & (u > 4) \end{cases} \\
\widetilde{g}(u) &= \begin{cases}
1 - \frac{9}{16} \sqrt{u} + \frac{1}{32} \sqrt{u^3} & (u \leq 4) \\
\frac{1}{\sqrt{u^3}} & (u > 4) \end{cases} \,,
\end{align}\end{subequations}
rather than $f(u)$ and $g(u)$ in equations~\eqref{eqn:diffeq} and~\eqref{eqn:ini-uss}.

Intuitively speaking, the mesoscopic particle moves in a time-dependent harmonic potential, whose coupling constant is tuned
in such a way that the potential energy of the particle in the trap matches the gravitational potential energy
between two particles, one sitting at the centre of the harmonic potential and the other one sitting at
$\sqrt{\langle r^2 \rangle}$.

From this intuitive picture it is already evident what could go wrong with such an approximation, if applied to the
self-gravitational potential in the Schr{\"o}\-din\-ger-New\-ton equation. For the case of a narrow wave-function, the gravitational potential
within the mesoscopic particle is, in fact, a harmonic potential, as is also indicated by the
approximation~\eqref{eqn:pot-narrow-sphere}.
However, in the wide wave-function case, where the mass density is well approximated by a delta distribution,
the gravitational potential for a Gaussian wave-function with $\erw{r^2} = u$ is
\begin{equation}
V_g(r) = -\frac{G\,m^2}{r} \, \mathrm{erf} \left(\sqrt{\frac{3r^2}{2u}}\right) \,.
\end{equation}
While this can be approximated by a quadratic function for small $r$, one obtains
$V_g \sim 1/r$ for large $r$.
Therefore, for the full Schr{\"o}\-din\-ger-New\-ton equation the part of the wave-function at
positions $r^2 \gg u$ will not see a quadratic potential, not even as an approximation, but a Coulomb-like
potential. Hence, from the evolution according to the full Schr{\"o}\-din\-ger-New\-ton equation one expects a weaker response to gravity than the
approximation by a harmonic potential according to equations~\eqref{eqn:sg-quad} and~\eqref{eqn:f-g-colin} suggests.

In Fig.~\ref{fig:plot-functions} the functions $f(u)$ and $\widetilde{f}(u)$ are plotted. There is a clear deviation
in the intermediate regime, where $u \approx R^2$. For wide wave-functions, the deviation decreases in magnitude. Nevertheless,
the functions also differ in their limiting behaviour for $u \to \infty$. While $f(u)$ approaches zero like $\sqrt{3/(16 \pi\,u^3)}$,
$\widetilde{f}(u)$ falls off like $1/\sqrt{4\,u^3}$.
Fig.~\ref{fig:plot-solution} shows how this can lead to a qualitatively different behaviour in the equations of motion.
The plotted solution corresponds to an Osmium sphere of about \unit{100}{\nano\meter} diameter. For a wide wave-function
both solutions show at least a similar tendency to slightly decrease in width and oscillate, even though with a significantly lower
frequency for the approximation with $f(u)$, $g(u)$ compared to $\widetilde{f}(u)$, $\widetilde{g}(u)$.
In the intermediate regime,
on the other hand, the solution according to eqs.~\eqref{eqn:diffeq} spreads only a little slower than the free solution,
while the solution according to eqs.~\eqref{eqn:sg-quad} and~\eqref{eqn:f-g-colin} is bound close to its initial width.

One can use the equation~\eqref{eqn:ini-uss} for the initial acceleration, in order to define a critical mass
(by the condition $\ddot{u}(0)=0$) which can serve as a rough estimate of whether the initial state will decrease or increase in width
upon free evolution. Comparison of the functions $g(u)$ and $\widetilde{g}(u)$ then shows that,
for wider wave-functions with $u \gtrsim (10R)^2$, the approximation chosen in ref.~\cite{Colin:2016,Colin:2014a}
underestimates this mass value by an approximate 20\,\%.

\section{Discussion}

We have shown how, under the assumption of a weak self-gravitational effect that leaves the shape of an initially
Gaussian wave packet intact, the free quantum evolution according to the Schr{\"o}\-din\-ger-New\-ton equation can be approximated by a
single parameter differential equation for the square of the wave-function width, $\erw{r^2}$.
The comparison with the previous approximation scheme used by Colin et~al.~\cite{Colin:2016,Colin:2014a}
for the very same purpose revealed deviations for wider wave-functions. While their results made use of an
approximation of the gravitational potential by a quasi-harmonic interaction, which was determined by a
rather ad hoc energy argument, here we provided a systematic derivation, based on the aforementioned approximation.

For the purpose of ref.~\cite{Colin:2016}, where self-gravitating effects were compared to decoherence
sources, both approximation schemes can serve as sufficiently good order of magnitude estimates.
The behaviour in Fig.~\ref{fig:plot-solution}b shows, however, that large deviations from the real
evolution according to the full equation are possible.

The alternative approximation scheme presented in this work is in no event inferior to the previously proposed
one by Colin et~al., as far as numerical efforts are concerned, since the functions $f(u)$ and $g(u)$ are known
analytically. This scheme can also be easily adapted to mass density distributions other than the homogeneous sphere.

\begin{acknowledgments}
I gratefully acknowledge funding by the John Templeton Foundation (Grant no. 39530) and the
German Research Foundation (DFG). I would like to thank Thomas Durt for helpful discussions.
\end{acknowledgments}

\appendix
\section{Derivation of \boldmath$\erw{V_g}(u)$ and \boldmath$\erw{\vec r \cdot \nabla V_g}(u)$}
We want to derive $\erw{V_g}$ with $V_g$ as defined in eq.~\eqref{eqn:sn-born-oppen}, for the homogeneous sphere
potential~\eqref{eqn:self-energy-homogeneous-sphere}. We write $u = \erw{r^2}$. The absolute value squared of the wave-function is
\begin{equation}
\abs{\psi}^2(u) = \left(\frac{3}{2\pi u}\right)^{3/2} \, \exp\left(-\frac{3 r^2}{2 u}\right) \,.
\end{equation}
First define $\alpha = 3 R^2 / (2 u)$ and
\begin{equation}
i(x) = \begin{cases}
\frac{6}{5}-2 x^2+\frac{3}{2} x^3-\frac{1}{5} x^5
& (x\leq 1)\,,\\
\frac{1}{2x}
& (x > 1)\,.
\end{cases}
\end{equation}
Then with appropriate substitutions we have
\begin{align}
\erw{V_g}(\alpha) &= -\frac{512 \,G\, m^2 \,\alpha^3}{\pi \,R} \int_0^\infty \D a \, \int_0^\infty \D b \, \int_{-1}^1 \D z
\nonumber\\
&\phantom{=}\; \times \mathrm{e}^{-4 \alpha (a^2 + b^2)} a^2 b^2 i(x) \\
x &= \sqrt{a^2 + b^2 - 2 a b z}
\end{align}
Further substituting $z$ by $x$, and finally $p = a+b$, $q = \abs{a-b}$, yields
\begin{align}
\erw{V_g}(\alpha) &= -\frac{512 \,G\, m^2 \,\alpha^3}{\pi \,R} \int_0^\infty \D a \, \int_0^\infty \D b \nonumber\\
&\phantom{=}\; \times \mathrm{e}^{-4 \alpha (a^2 + b^2)} a \, b \, \int_{\abs{a-b}}^{a+b} \D x \, x \, i(x) \nonumber\\
&= -\frac{128 \,G\, m^2 \,\alpha^3}{\pi \,R} \int_0^\infty \D p \, \int_0^p \D q \nonumber\\
&\phantom{=}\; \times \mathrm{e}^{-2 \alpha (p^2 + q^2)} (p^2 - q^2) \, \left(J(p) - J(q)\right)
\end{align}
with
\begin{equation}\vspace{-2pt}
J(x) = \begin{cases}
\frac{3}{5}x^2-\frac{1}{2}x^4+\frac{3}{10}x^5-\frac{1}{35}x^7
& (x\leq 1)\,,\\
\frac{1}{2}x
& (x > 1)\,.
\end{cases}
\end{equation}
In order to obtain $\erw{\vec r \cdot \nabla V_g}(u)$ we have to repeat essentially the same calculation with the
replacement
\begin{equation}
i(x) \to (a^2 - b^2 + x^2) \frac{i'(x)}{2x} \,,
\end{equation}
where the $a^2$ and $b^2$ terms cancel for symmetry reasons.
This yields the same final relation
\begin{multline}
\erw{\vec r \cdot \nabla V_g}(\alpha) = -\frac{128 \,G\, m^2 \,\alpha^3}{\pi \,R} \int_0^\infty \D p \, \int_0^p \D q \, \\
\times \mathrm{e}^{-2 \alpha (p^2 + q^2)} (p^2 - q^2) \, \left(K(p) - K(q)\right)
\end{multline}
but with
\begin{equation}
K(x) = \begin{cases}
-\frac{1}{2}x^4+\frac{9}{20}x^5-\frac{1}{14}x^7
& (x\leq 1)\,,\\
-\frac{1}{4}x
& (x > 1)\,.
\end{cases}
\end{equation}

\end{document}